\documentstyle[aps,preprint,tighten,floats,epsfig]{revtex}

\begin{document}

\draft
\preprint{
\vbox{
\hbox{January 2003}
\hbox{ADP-02-79/T518}
\hbox{JLAB-THY-02-32}
}}

\title{Nuclear Shadowing at Low $Q^2$}

\author{W.~Melnitchouk}
\address{Jefferson Lab,
	12000 Jefferson Avenue,
	Newport News, VA 23606}
\author{A.~W.~Thomas}
\address{Special Research Centre for the Subatomic Structure of Matter,\\ 
	and Department of Physics and Mathematical Physics, \\ 
	University of Adelaide, 5005, Australia}

\maketitle

\begin{abstract}
We re-examine the role of vector meson dominance in nuclear shadowing
at low $Q^2$.
We find that models which incorporate both vector meson and partonic
mechanisms are consistent with both the magnitude and the $Q^2$ slope of
the shadowing data.
\end{abstract}

\vspace*{0.5cm}

There has been renewed interest recently in the problem of nuclear
shadowing in structure functions at low and intermediate $Q^2$.
In part, this has been prompted by the analysis of the NuTeV Collaboration
\cite{NUTEV} of neutrino--nucleus cross sections and subsequent questions
about nuclear shadowing corrections when extracting nucleon quark
distributions or electroweak parameters \cite{MILLER,KSY,KUMANO}.
Indeed, shadowing in neutrino scattering has received considerably less
attention than in electromagnetic reactions, and currently there are 
proposals to utilize high intensity neutrino and antineutrino beams to
perform high statistics measurements of $\nu/\bar\nu$--nucleus cross
sections at Fermilab \cite{NUMI}.
A pressing need exists, therefore, to understand the differences between
nuclear shadowing effects in charged lepton and neutrino scattering
\cite{BOROS,KULAGIN}, especially at low $Q^2$.

An extensive review of both data and models of nuclear shadowing was
given recently by Piller and Weise \cite{PW}.
Before one can reliably tackle nuclear corrections in neutrino
scattering, however, it is vital to determine the relevant degrees of
freedom responsible for shadowing in charged lepton scattering, where
data are much more copious.
The best available data on nuclear shadowing, including the $Q^2$
dependence, are from the New Muon Collaboration \cite{NMC_C,NMC_D,NMC_SN}.
We shall concentrate on a model based on a two-phase picture of nuclear
shadowing \cite{MT_D,MT_A,MT95}, similar to that pioneered by Kwiecinski
and Badelek \cite{KB,BK,KW}, which we published just before the release of
the final NMC data \cite{NMC_SN}.
For clarity we briefly review this model.

At high virtuality the interaction of a photon with a nucleus can be
efficiently parameterized through a partonic mechanism, involving
diffractive scattering through the double and triple Pomeron \cite{GOUL}.
{}For $Q^2 \agt 2$~GeV$^2$, the contribution to the nuclear structure
function $F_2^A$ (per nucleon) from this mechanism can be written as
\begin{eqnarray}
\label{dFAP}
\delta^{({\rm I\!P})} F_2^A(x,Q^2)
&=& {1 \over A} \int_{y_{min}}^A\ dy\
f_{{\rm I\!P}/A}(y)\ F_2^{{\rm I\!P}}(x_{{\rm I\!P}},Q^2)\ ,
\end{eqnarray}
where $f_{{\rm I\!P}/A}(y)$ is the Pomeron (${\rm I\!P}$) flux, and
$F_2^{\rm I\!P}$ is the effective Pomeron structure function \cite{IS}.
The variable $y = x (1+M_X^2/Q^2)$ is the light-cone momentum fraction
carried by the Pomeron ($M_X$ is the mass of the diffractive hadronic
debris), and $x_{\rm I\!P} = x/y$ is the momentum fraction of the
Pomeron carried by the struck quark.
The dependence of $F_2^{\rm I\!P}$ on $Q^2$ at large $Q^2$, in the
region where perturbative QCD can be applied, arises from radiative
corrections to the parton distributions in the Pomeron \cite{KW,CHPWW},
which leads to a weak, logarithmic, $Q^2$ dependence for the shadowing
correction $\delta^{({\rm I\!P})} F_2^A$.
Alone, the ${\rm I\!P}$ contribution to shadowing would give a
structure function ratio $F_2^A/F_2^D$ that would be almost flat for
$Q^2 \agt 2$~GeV$^2$ \cite{CAPELLA}.

On the other hand, the description of shadowing at low $Q^2$ requires a
higher-twist mechanism, such as vector meson dominance (VMD), which can
map smoothly onto the photoproduction limit at $Q^2=0$.
The VMD model is empirically based on the observation that some aspects
of the interaction of photons with hadronic systems resemble purely
hadronic interactions \cite{BAUER,SCHSJ}.
In QCD language this is understood in terms of the coupling of the photon
to a correlated $q\bar q$ pair with low invariant mass, which may be
approximated as a virtual vector meson.
One can then estimate the amount of shadowing in terms of the multiple
scattering of the vector meson using Glauber theory \cite{GLAUBER}.
The corresponding VMD correction to $F_2^A$ is
\begin{eqnarray}
\label{dFAV}
\delta^{(V)} F_2^A(x,Q^2)
&=& {1 \over A} { Q^2 \over \pi }
\sum_V { M_V^4\ \delta\sigma_{VA} \over f_V^2 (Q^2 + M_V^2)^2 }\ ,
\end{eqnarray}
where $\delta \sigma_{VA}$ is the shadowing correction to the vector
meson--nucleus cross section, $f_V$ is the photon--vector meson coupling
strength \cite{BAUER}, and $M_V$ is the vector meson mass.
In practice, only the lowest mass vector mesons
($V = \rho^0, \omega, \phi$) are important at low $Q^2$.
(Inclusion of higher mass states, including continuum contributions,
leads to so-called generalized vector meson dominance models \cite{GVMD}.)
The vector meson propagators in Eq.~(\ref{dFAV}) lead to a strong $Q^2$
dependence of $\delta^{(V)} F_2^A$ at low $Q^2$, which peaks at
$Q^2 \sim 1$~GeV$^2$, although one should note that the nucleon structure
function itself also varies rapidly with $Q^2$ in this region.
For $Q^2 \rightarrow 0$ and fixed $x$, $\delta^{(V)} F_2^A$ disappears 
because of the vanishing of the total $F_2^A$.
Furthermore, since this is a higher twist effect, shadowing in the VMD
model dies off quite rapidly between $Q^2 \sim 1$ and 10 GeV$^2$, so that
for $Q^2 \agt 10$ GeV$^2$ it is almost negligible --- leaving only the
diffractive partonic term, $\delta^{({\rm I\!P})} F_2^A$.

The accuracy of the model can be tested by looking for deviations from
logarithmic $Q^2$ dependence of shadowing at low and intermediate $Q^2$.
Actually, a detailed analysis of the $Q^2$ dependence of the NMC data,
as well as the lower-$Q^2$ Fermilab-E665 data \cite{E665}, was performed
in Refs.\cite{MT_A,MT95} for various nuclei from $A=2$ to $A=208$
({\em viz.}, for D, Li, Be, C, Al, Ca, Fe, Sn, Xe and Pb).
Ratios of $F_2^A/F_2^D$ were calculated \cite{MT_A,MT95} for a range of
$x$ ($10^{-5} \alt x \alt 0.1$) and
$Q^2$ ($0.03 \alt Q^2 \alt 100$~GeV$^2$).
Subsequent to these analyses, high precision data on the $Q^2$ dependence
of Sn/C structure function ratios were published \cite{NMC_SN}, which
provided the first detailed evidence concerning the $Q^2$-dependence of
nuclear shadowing.

\begin{figure}[t]
\begin{center}
\epsfysize=8cm
\leavevmode
\epsfbox{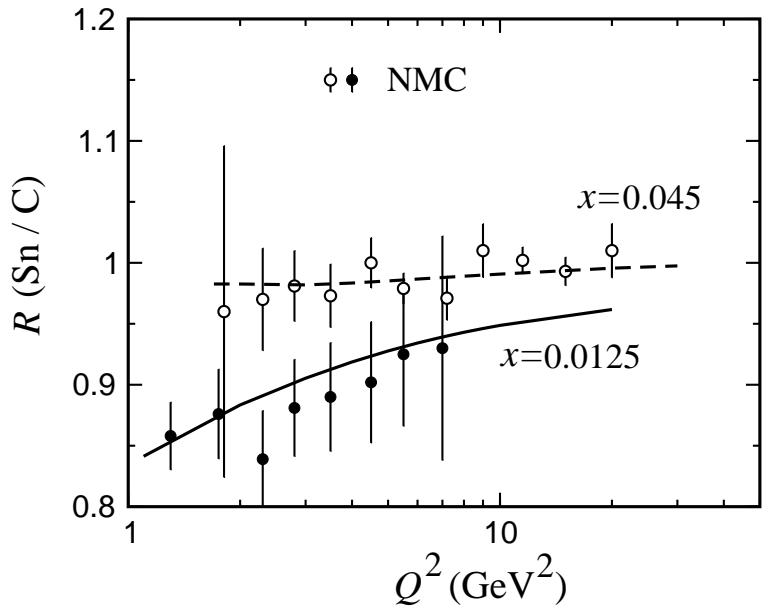}
\vspace*{0.5cm}
\caption{$Q^2$ variation of the Sn/C structure function ratio in
	the model of Ref.~\protect\cite{MT95} for $x=0.0125$ (solid)
	and $x=0.045$ (dashed).
	The data are from NMC \protect\cite{NMC_SN}, with statistical
	and systematic errors added in quadrature.}
\end{center}
\end{figure}

In Fig.~1 we show the calculated ratio
$R ({\rm Sn/C}) \equiv F_2^{\rm Sn}/F_2^{\rm C}$ as a function of $Q^2$
for $x=0.0125$ (solid curve) and $x = 0.045$ (dashed), compared with the
NMC data \cite{NMC_SN}.
The overall agreement between the model and the data is clearly excellent.
In particular, the observed $Q^2$ dependence of the ratios is certainly
compatible with that indicated by the NMC data.
At large $Q^2$ ($Q^2 \agt 10$ GeV$^2$) the $Q^2$ dependence is very weak,
as expected from a partonic, leading-twist mechanism \cite{MT95} ---
see also Refs.\cite{QIU,BL,NZ,KUM,KOP}.
In the smallest $x$ bins, however, the $Q^2$ values reach down to
$Q^2 \approx 1$~GeV$^2$.
The data on the C/D and Ca/D ratios analyzed in Ref.~\cite{MT95} at even
smaller $x$ ($x \agt 0.0003$) extend down to $Q^2 \approx 0.05$~GeV$^2$.
This region is clearly inaccessible to any model involving only a
partonic mechanism, and it is essential to invoke a non-scaling mechanism
here, such as vector meson dominance.
One should also note that, even though the shadowing corrections may
depend strongly on $Q^2$, because the nucleon structure function itself
is rapidly varying at low $Q^2$, the $Q^2$ dependence of the ratio will
not be as strong as in the absolute structure functions.
In any case, the fact that the two-phase model \cite{MT95} describes the
NMC data over such a wide range of $Q^2$ gives one added confidence in
extending this model to neutrino scattering \cite{BOROS}.

To illustrate the $Q^2$ dependence of $R$ over the full range of $x$
covered in the NMC experiment, Arneodo {\em et al.} \cite{NMC_SN}
parameterized the Sn/C ratio as $R({\rm Sn/C}) = a + b \ln Q^2$, and
extracted the logarithmic slopes $b=dR/d\ln Q^2$ as a function of $x$.
%
%
As illustrated in Fig.~2, the NMC find that the slopes are positive and
differ significantly from zero for $0.01 < x < 0.05$, indicating that the
amount of shadowing decreases with increasing $Q^2$ \cite{NMC_SN}.
The logarithmic slope $b$ is found to decrease from $\approx 0.04$ at the
smallest $x$ value to zero at $x \agt 0.06$.
The result of the model calculation \cite{MT95} is perfectly consistent
with the NMC data over the full range of $x$ covered, as Fig.~2
demonstrates (see also Fig.~3(b) of Ref.~\cite{MT95}).
In particular, the ${\rm I\!P}$-exchange mechanism alone, modified
by applying a factor $Q^2/(Q^2 + Q_0^2)$ \cite{BK,DLQ} to ensure that
$\delta^{({\rm I\!P})} F_2^A \rightarrow 0$ as $Q^2 \rightarrow 0$,
is clearly insufficient \cite{CAPELLA} to describe the logarithmic slope
in $Q^2$ at low $x$, whereas the addition of a VMD component 
does allow one to describe the data quite well 
(the shaded region indicates an estimate of the uncertainty in the model
calculation).
\begin{figure}[t]
\begin{center}
\epsfysize=8cm
\leavevmode
\epsfbox{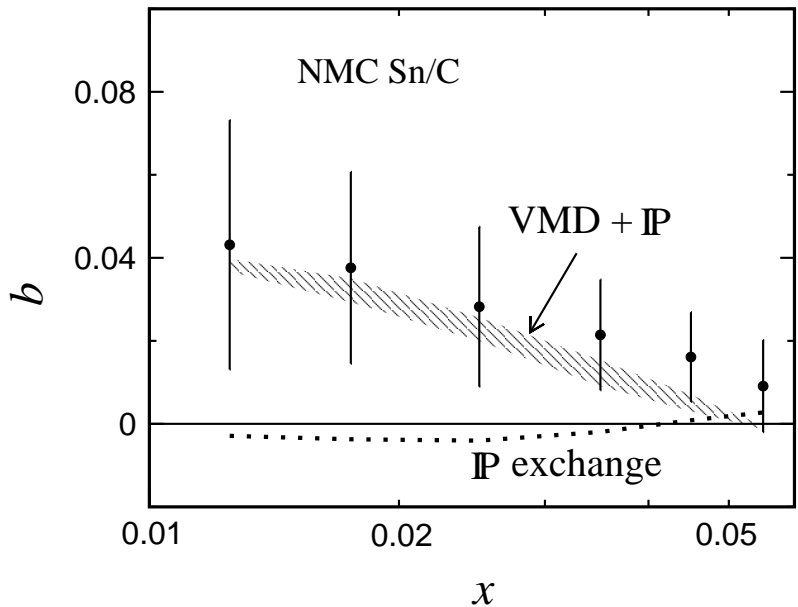}
\vspace*{0.5cm}
\caption{Logarithmic slope, $b$, in $Q^2$ of the NMC Sn/C ratio as a
	function of $x$ \protect\cite{NMC_SN}, compared with the nuclear
	shadowing model of Ref.~\protect\cite{MT95}.
	The statistical and systematic errors are added in quadrature.}
\end{center}
\end{figure}

In summary, the results of this analysis demonstrate that a combination
of VMD at low $Q^2$ to describe the transition to the photoproduction
region, with parton recombination, parameterized via ${\rm I\!P}$-exchange,
at high $Q^2$ allows one to accurately describe shadowing in
electromagnetic nuclear structure functions over a large range of $Q^2$.
As well as confirming that higher twist effects are numerically important
at intermediate $Q^2 \sim 1$--4~GeV$^2$, our findings also suggest that the
two-phase model can serve as an excellent basis on which to reliably
tackle the question of shadowing in neutrino reactions.

\acknowledgements

We thank M.~Arneodo, A.~Br\"ull, and M.~Szleper for providing the NMC
data.
This work was supported by the Australian Research Council, and the
U.S. Department of Energy contract \mbox{DE-AC05-84ER40150}, under which
the Southeastern Universities Research Association (SURA) operates the
Thomas Jefferson National Accelerator Facility (Jefferson Lab).


\end{document}